# COMPETENCE BUILDING FRAMEWORK REQUIREMENTS FOR INFORMATION TECHNOLOGY FOR EDUCATIONAL MANAGEMENT

Rakesh Mohan Bhatt

Department of Computer Science and Engineering,
HNB Garhwal University, Srinagar Uttrakhand, India
rmbhatt77@yahoo.com

## ABSTRACT

*Progressive efforts have been evolving continuously for the betterment of the services of the Information Technology for Educational Management(ITEM). These services require data intensive and communication intensive applications. Due to the massive growth of information, situation becomes difficult to manage these services. Here the role of the Information and Communication Technology (ICT) infrastructure particularly data centre with communication components becomes important to facilitate these services. The present paper discusses the related issues such as competent staff, appropriate ICT infrastructure, ICT acceptance level etc. required for ITEM competence building framework considering the earlier approach for core competences for ITEM. It this connection, it is also necessary to consider the procurement of standard and appropriate ICT facilities. This will help in the integration of these facilities for the future expansion. This will also enable to create and foresee the impact of the pairing the management with information, technology, and education components individually and combined. These efforts will establish a strong coupling between the ITEM activities and resource management for effective implementation of the framework.*

## 1. INTRODUCTION

We have seen the strength of Information Technology (IT) in providing innovation and competitive edge to any organization (Talero & Gandette 1995). Further, using the Management Information System (MIS), many benefits have been observed in the education system (Visscher et.al. 2001) ,that is how the MIS systems affect the control of educational institution and educators performing administrative functions too. The use of educational information management system provides a greater degree of standardization of administrative functions. Embedded software enables a kind of automation to some extent the job to be performed without any control. But, policies of decentralization are yet to be achieved. Three levels of training (Donnelly, 2000) have been claimed as appropriate to both the leadership team and administrative staff viz. training in generic software, training in the use of institution specific MIS software, and training in the use of the internet. Additionally, the leadership team should be trained to use data or information for the improvements of the educational standards. So, ICT acceptance level should be high among the members.

During the ITEM-2002, a Discussion Group was formed. This discussion group analyzed that the emerging technologies are not very well adopted by the educational institutions. These institutions are failed to provide the coherent and effective training programs as reported the unsatisfactory use of ICT means for the teaching, learning and administrative purposes (Lambert & Nolan 2002, DfES 2002, Newton 2002). Therefore, appropriate ICT infrastructure should be considered to facilitate the services required. As a result, the Discussion Group designed a model or framework to enable to plan the ITEM training and achieve an ITEM- competent staff ( Ian Selwood & et. al. 2003). The present paper analyzes the appropriate ICT infrastructure, competent staff, and ICT acceptance level required in the following sections of ICT infrastructure and ITEM competence building framework for its proper implementation.

## 2. ICT INFRASTRUCTURE

Information management system mend for the educational institute enables central education authorities to exercise a form of "control at a distance" over the institutional operation without appearing to intervene directly. Enactment of more e-enabled services learning activities through asynchronous or synchronous processes is being effectively performed.

While analyzing the adoptability of emerging technologies, the designed framework by the Discussion Group is consisted 36 competences across the three dimensions. The first dimension points for the four inextricably linked dimensions; they are Information, Technology, Education and Management. The second dimension is set for the management and planning concerned. For this purpose, it has three levels viz. operational, tactical and strategic. The last, third dimension defines the stage of growth. Three growth labels are considered i.e. initiation, expansion and embedded. Thus, altogether this form a matrix of 4 * 3 * 3 and out of this, 36 competences/activities have been emerged out.

We cannot limit the workability in the collaborative environment and this phenomenon has been increased tremendously with the effective use of IT. For effective global operations, use of IT is fundamental (King & Sethi 1999) as it coordinates the dispersed activities and establish coalition in between different activities. So, organizations are trying to uncover this potential of usage of IT which could only be achieved if the organization has the appropriate technological infrastructure. For example, issues as proposed (Bhatt, 2007) such as wireless technology, cabling, Ethernet & Asynchronous Transfer Mode (ATM), iSCSI and IPv6 are important to consider to support and integrate e-communication for the massive information. In the recent past, progressive efforts have been evolving continuously for the betterment of the services of the information system for educational management, which require data intensive and communication intensive applications. Due to the massive growth of information, situation becomes difficult to manage these services and in this respect the role of ICT infrastructure particularly data centre with communication components becomes important to facilitate these services. Therefore, the emerging concept of data centre model (Clabby Analytics, 2008) can be considered to handle data intensive and data communication intensive applications becomes inevitable for the e-learning process. Till data modernization of data centre have been performed in the business organization (Barnett, 2008). But a lot of transformation is needed in the educational institutions. This is because modern data centre are based on object-oriented technology, standard format for data sharing e.g. XML; Java technologies & internet protocol (IP) for interoperability and network communication. During the data intensive applications, large data sets are connected for distribution and may be further connected with intranet, extranet and internet. For this purpose, data centre provides data searching, retrieval and sharing quickly via its data consolidation process. For communication intensive applications, high efficient data search, retrieval and large scale collaborative multimedia system supports are required. For example, real-time multimedia interactive requires high network bandwidth for efficient data transmission in the collaborative working environment. This is achieved with the help of server virtualization process.

Therefore, through consolidation and virtualization, performance of systems and communication network for storage and sharing resources is enhanced for data as well communication intensive applications where increased traffic poses bandwidth constraints. So, to enable network responsible (McGillicuddy, 2012), server virtualization is essential to ease the operation of data centre infrastructure.

In addition to the benefits in providing standardization of administrative functions as discussed above, another kind of standardization should be considered for technological advancement so as to integrate the existing technological infrastructure components with their counterparts to come in future i.e. components should be compatible in their operations with each other. This technological integration would automatically integrate the collaborative and sharing efforts at the interaction level of the defined three axis of the model so that mapping with its existing policies and programs and investigate goodness of fit in accordance to the ITEM competencies can be achieved. In order to control the growth, the feedback process at the management level can be enforced.

## 3. COMPETENCE BUILDING FRAMEWORK FOR ITEM

Visscher and Branderhost (2001) have addressed five skill areas - to recognize the information value & policy development, to determine the type of information needed, to discover the information out of the MIS, to interpret information from the MIS, and to make decision and evaluate the policy. But, in addition to these five skills, one more skill is required. It can be named as the feedback-skill and is useful while interacting with the existing MIS at any level which would strengthen the management for its strategic, tactical and operational competences. These levels can be categorized as level-1, level-2 and level-3. The type of levels can be defined as the Level-1 is the type of structured competence; which is a prerequisite for transition to the Level-2, a semi-structured competence and Level-2 competence is a prerequisite for the transition to the Level-3, which is a kind of unstructured competence. So, the order of priority among the competences to be given is a complex task and pertains to the strategic competence; which comes at the top of all these three levels, partly can be matched with the Level-3 competence.

Further, where to initiate or expand or embed in order to retain the balance as such and for how much time. For example, strategic competence takes less time to take quick decisions; technical competence takes some more time and operate competence takes longer time consumed in providing its services and interactions processes.

Thus, to evaluate the ICT skills and pedagogy for ICT enabled teaching process, following points are needed to be analyzed:

a) The level of acceptance of ICT skill varies from school level to post-graduate level of institutions.

b) The level of ICT skill also reflects the level of procurement and integration of ICT equipments.

c) ICT skill is also influenced by the availability of the ICT facilities exist in the institute.

These points should be considered the most important activity, hence the management team/authority (comprising educationists and technocrats) should have a very broad vision so as to open ICT windows at all levels of learning and teaching to enhance the availability of the ICT usage. This phenomenon will also empower the management process through the feed back received instantly by building the ICT means besides the ICT skill. However, this empowerment will also be benefited by the advantages from the proposed model (Ian Selwood et.al., 2003) such as it is platform independent, and it is descriptive and prescriptive. In meeting the three-group competencies such as using tools interactively, interacting in heterogeneous groups and acting autonomously, teaching process should also be under the adequate educational environment(Kollee, C. et. al., 2009). Therefore, suitable mapping can be done with the existing activities for its fitness.

The above analysis covering all the three points would help the strategic support to distribute the ICT resources uniformly. Essentially a problem of management, relates with the process of managing the information systems can be implemented by providing best support technologies components to maintain the richness of the information. So, this will ease the management work in the proposed model of ITEM competence by considering the relevance and compatible technology for education and vis-à-vis segregate them ( information, technology and education) for strategic, tactical and operational levels of actions. Then the steps towards the initiation, expansion and embedment can be taken accordingly.

This would lead to the ITEM competence building properly. Therefore, it is recommended for the management process to analyze the levels of:

a) acceptance of ICT skill,

b) procurement and integration of ICT tools/ equipments, and

c) ICT facilities,

This analysis will help to create and foresee the impact of the pairing the management with information, technology, and education components individually and combined. This would also improve the base for the ITEM training and achieve an ITEM-competent technology and staff. This will then become easy to anticipate the inevitable changes. Attributed benefits accruing out of it can also be traced and understood.

To extract standards to support teaching and learning for ITEM "Technological Standards for School Administrators"(TSSA collaborative 2001) is suitable approach. Further, Bhatt (2007) has pointed out that the ITEM-competence is also strongly influenced by the appropriate selection and use of information and technological components. Therefore, this kind of integration, data centre concept and the TSSA approach can be considered for the better ITEM competence building. Further, the UNESCO ICT Competency Framework (Wallet, 2014) has also been designed for Teachers which is a useful tool to inform education policymakers, educators and providers of professional learning of the role of ICT in educational reform. It also assists Member States in developing national ICT competency standards. Emphasis has been given on to collaborate in problem-solving and creative learning so as to enhance the student outcomes. It this connection, it is also necessary to consider the procurement of standard and appropriate ICT facilities. This will help in the integration of these facilities for the future expansion. This will also enable to create and foresee the impact of the pairing the management with information, technology, and education components individually and combined. These efforts will establish a strong coupling between the ITEM activities and resource management for effective implementation of the framework.

## 4. CONCLUSIONS

Legacy storage and communication support systems should be transformed to modernize the data centre in order to resolve constantly changing demands for various kinds of applications overcoming the interoperability, data sharing problems and quality of information delivery. The aforesaid efforts could provide the benefits in the proposed structured approach of the model by solving the complex, dynamic and distributed behavior of the information operations with the help of virtualization, data consolidation and network management. Therefore, for the better competences and to accelerate the interactions more effectively, it is strongly believed that with appropriate implementation of technological components and analyses of management with information, technology, and education components individually and combined; a strong coupling can be established between the ITEM activities and resource management to nurture the ITEM building process.

## ACKNOWLEDGEMENTS

The author is sincerely thankful to HNB Garhwal University, Srinagar and Uttrakhand State Council for Science & Technology for providing financial support to present this paper.

## AUTHORS

**Dr  Rakesh Mohan Bhatt**

Dr R M Bhatt has experience of about 30 years in the field of Computer Applications. He is associated with the IFIP WG on AI, and Education. He is also a senior life member of Computer Society of India. He has been Visiting Professor at SS Cyril University, Slovakia and recipient of Commonwealth Fellowship Award for Academic Exchange. He has published around 45 papers.

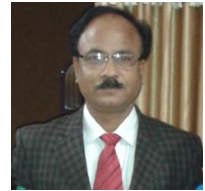